\documentclass[aip,jcp,reprint,twocolumn,showkeys]{revtex4-1}
\pdfoutput=1 

\usepackage[T1]{fontenc}
\usepackage[utf8]{inputenc}
\usepackage{color}
\usepackage{graphicx}
\usepackage{multirow}
\usepackage{amsmath,amssymb}
\usepackage{calc}
\usepackage{feynmp}
\usepackage{subfigure}
\usepackage[inline]{asymptote}
\usepackage[english]{babel}
\begin{asydef}
  import graph;
  defaultpen(fontsize(11));
  real goldenRatio = (1+sqrt(5))/2;
  void sizeRatio(
    picture pic=currentpicture,
    real width=0, real height=0, real ratio=goldenRatio
  ) {
    if (width > 0) {
      size(pic, width, width/ratio, IgnoreAspect);
    } else {
      size(pic, height*ratio, height, IgnoreAspect);
    }
  }
  void plotXY(
    picture pic=currentpicture,
    real[][] data,
    pen p=defaultpen,
    Label l="",
    marker m=nomarker
  ) {
    data = transpose(data);
    draw(pic, graph(pic,data[0],data[1]), p, l, m);
  }
  void plotXYDY(
    picture pic=currentpicture,
    real[][] data,
    pen p=defaultpen,
    Label l="",
    marker m=nomarker
  ) {
    data = transpose(data);
    real[] zeros = array(data[0].length, 0.0);
    errorbars(pic, data[0], data[1], zeros, data[2], p);
    draw(pic, graph(data[0],data[1]), p, l, m);
  }
  void axisXY(
    picture pic=currentpicture,
    Label xLabel="", Label yLabel="",
    pair min=(-infinity,-infinity), pair max=(+infinity,infinity),
    pen p=currentpen,
    ticks xTicks=LeftTicks, ticks yTicks=RightTicks
  ) {
    limits(pic, min, max, crop=true);
    xaxis(pic, xLabel, BottomTop, min.x, max.x, xTicks);
    yaxis(pic, yLabel, LeftRight, min.y, max.y, yTicks);
  }
  void axisY(
    picture pic=currentpicture,
    Label l="",
    real min=-infinity, real max=+infinity,
    pen p=currentpen,
    ticks ticks=RightTicks
  ) {
    ylimits(pic, min, max, crop=true);
    yaxis(pic, LeftRight, min, max, ticks);
  }
\end{asydef}

\newcommand{\eps}{\varepsilon}

\newcommand{\diagramBox}[2][0.5]{\raisebox{0.5ex-#1\height}{#2}}

\begin{document}

\title{
  Low rank factorization of the Coulomb integrals
  for periodic coupled cluster theory
 }
\author{Felix Hummel}
\email{f.hummel@fkf.mpg.de}
\affiliation{Max Planck Institute for Solid State Research, Heisenbergstra{\ss}e 1, D-70569 Stuttgart, Germany}
\author{Theodoros Tsatsoulis}
\affiliation{Max Planck Institute for Solid State Research, Heisenbergstra{\ss}e 1, D-70569 Stuttgart, Germany}
\author{Andreas Gr\"uneis}
\email{a.grueneis@fkf.mpg.de}
\affiliation{Max Planck Institute for Solid State Research, Heisenbergstra{\ss}e 1, D-70569 Stuttgart, Germany}
\affiliation{
  Department Chemie, Technische Universität München (TUM)
  Lichtenbergstra{\ss}e 4, D-85747 Garching, Germany
}

\begin{abstract}
We study the decomposition of the Coulomb integrals of periodic systems
into a tensor contraction of six matrices of which only two are distinct.
We find that the Coulomb integrals
can be well approximated in this form already with small matrices
compared to the number of real space grid points.
The cost of computing the matrices scales as $\mathcal O(N^4)$ using a regularized form
of the alternating least squares algorithm.
The studied factorization of the Coulomb integrals can be exploited
to reduce the scaling of the computational cost of expensive tensor contractions
appearing in the amplitude equations of coupled cluster methods with respect to system size.
We apply the developed methodologies to calculate the adsorption energy of a single water molecule on
a hexagonal boron nitride monolayer in a plane wave basis set and periodic boundary conditions.

\keywords{coupled cluster; Coulomb integrals; tensor rank decomposition; canonical polyadic decomposition}
\end{abstract}

\maketitle

\section{Introduction}

The high dimensionality of the many electron wave function is one of the most limiting
factors in applying highly accurate electronic structure theories to the solution of the many electron
Schr\"odinger equation for real materials on an {\it ab initio} level~\cite{szabo_modern_1996}.
Many of the most widely used wave function based
theories have a good balance between accuracy, computational cost and the number of parameters used to
approximate the exact wave function.
The efficiency is strongly affected by the computational complexity required to evaluate resultant expectation values.

Tensor rank decompositions (TRD) and low rank tensor approximations are ubiquitous in the
field of electronic structure theory calculations.
These techniques are essential to reduce the computational cost and memory footprints
to calculate and store the approximate many electron wave function.
Already the simplest level of approximation, Hartree--Fock (HF) theory can be
regarded as a low rank approximation,
employing an antisymmetrized outer product of one electron orbitals
to approximate the full wave function.
This low rank tensor approximation is identical to a single Slater determinant.
However, HF theory neglects electronic correlation effects.
Electronic correlation effects can be captured by extending the wave function basis
with additional determinants. For this purpose, excited HF determinants can be employed.
They are constructed by replacing occupied orbitals with
unoccupied orbitals, forming a complete and orthogonal basis.
Computationally, the basis of (excited) Slater determinants is very convenient. It introduces
a large degree of sparsity to the full many electron Hamiltonian and  simplifies the solution
of the many electron problem.
Most entries in the sparse many electron Hamiltonian can be calculated directly from  electron repulsion integrals.
The memory footprint for the storage of these integrals in a canonical basis is very large and grows rapidly
with respect to the number of orbitals.
Therefore it is often necessary to calculate these integrals in a computationally efficient on the fly manner.
The most widely used schemes for the calculation of electron repulsion integrals include~\cite{Reine2012,Hellweg2007,Ahlrichs2004,Weigend2005,Manby2001}:
(i) prior calculation of the integrals in the employed atomic orbital basis and its subsequent
transformation into a molecular orbital basis, and (ii) employing the resolution of identity approach.
Computationally the resolution of identity approach is more efficient because it requires the calculation
and storage of intermediate quantities with at most three indices. In passing we note that the expression of the integrals
in terms of these intermediate quantities allows for rearranging nested summations in ring coupled cluster theories
such that the scaling of the computational cost with respect to the system size
can be reduced~\cite{Gruneis2009}.

Coupled cluster theory can also be viewed as a low rank tensor approximation to the exact
configuration interaction wave function coefficients in the Slater determinant
basis. The exponential ansatz used in coupled cluster theories effectively
approximates the coefficients of highly excited determinants by outer products of cluster
amplitudes with a lower rank.
However, increasingly accurate levels of coupled cluster theories lead to increasingly steep polynomial
scalings of the computational cost and  memory with respect to the studied system sizes.
In this work we seek to reduce the computational cost of coupled cluster theories without introducing
additional approximations on the level of the employed wave function.
This can be achieved by employing low rank tensor approximation techniques for the decomposition of the
two electron integrals and the corresponding intermediate quantities obtained from the resolution of identity approach.
Using the low rank decomposition, the nested summations in the amplitude equations of
distinguishable cluster singles and doubles theory
(DCSD)~\cite{kats_distinguishable_2016,kats_communication:_2013,kats_accurate_2015},
as well as of linearized coupled cluster singles and doubles theory 
can be rearranged such that the scaling of the computational cost is reduced
from $\mathcal O(N^6)$  to $\mathcal O(N^5)$ without any further
approximations. Additionally, we highlight that the low rank factorization of
the Coulomb integrals also allows for reducing the scaling of the
computationally most expensive terms in coupled cluster singles and doubles
(CCSD) theory using a plane wave basis set.

We note that the methods outlined in this work share many similarities with other approaches that
aim at the reduction of the computational cost in correlated wave function based theories.
In particular we want to point out that the tensor hypercontraction (THC) technique introduced
by Hohenstein \emph{et al.} in Refs.~\onlinecite{hohenstein_tensor_2012,parrish_tensor_2012,hohenstein_communication:_2012}
also performs a low rank tensor decomposition of the Coulomb integrals.
Furthermore, a similar approach to the tensor rank decomposition method introduced in this work was discussed by
Shenvi \emph{et al.} in Ref.~\onlinecite{shenvi_tensor_2013}.
In the work of Benedikt \emph{et al.} it was shown that tensor rank decomposition techniques
can even be applied to the decomposition of the coupled cluster amplitudes directly~\cite{benedikt_tensor_2013}.
However, in contrast to the methods mentioned above we introduce an efficient numerical procedure to achieve the low rank
tensor decomposition of the Coulomb integrals in periodic systems without the necessity of defining an {\it a priori} real
space grid as it is the case for THC methods.

\subsection{Structure of this work}
The factorization of the Coulomb integrals tensor is obtained in two steps.
In Section \ref{sec:OAF} we first discuss how the Coulomb integrals
can be decomposed into a contraction of two third order tensors:
$V^{pq}_{sr} \approx {\Gamma^\ast}^{pF}_s \Gamma^q_{rF}$,
where we refer to $\Gamma^q_{rF}$ as optimized Coulomb vertex.
Subsequently, we perform a Tensor Rank Decomposition (TRD) of
the optimized Coulomb vertex into a contraction of three matrices:
$\Gamma^q_{rF} \approx \Lambda^R_F{\Pi^\ast}^q_R\Pi^R_r$.
Section \ref{sec:TRD} describes the employed algorithms to compute this
factorization.

Section \ref{sec:Application} outlines
how this factorization can be employed by quantum chemistry methods 
and in Section \ref{sec:Results} we study the application
of the discussed approximations to different systems.
Subsection \ref{ssc:LiH} focuses on the convergence of the TRD
for total energies of the LiH solid, while we
compute coupled cluster adsorption energies of water on
the surface of a single BN sheet
in Subsection \ref{ssc:Adsorptions}.

\subsection{Notation}
We imply a sum over all free indices occurring at least twice within a product but nowhere else.
We will use the letters $i,j,k,l$ to label occupied spin orbitals, $a,b,c,d$ to
label virtual spin orbitals and $p,q,r,s$ to label general spin orbitals.
The letters $R,S,T,U$ are used to denote elements of the rank decomposition.
The conjugate transpose of a tensor such as $A^q_r$ is denoted by ${A^\ast}^r_q$
where lower and upper indices are swapped.
Sequence numbers in iterations are given in superscript within parentheses,
as in $A^{(n)}$.
The Frobenius norm of a tensor $A$ is denoted by $\left\Vert A\right\Vert$.
Examples are:
\begin{align}
  V^{ab}_{ij}V^{ij}_{ab}
  \quad =& \quad
    \sum_{a,b\in{\rm virt.},i,j\in{\rm occ.}} V^{ab}_{ij}V^{ij}_{ab}\\
  T_{ijk} - A_{iR}B_{jR}C_{kR}
  \quad =& \quad
    T_{ijk} - \sum_{R=1}^{N_R}A_{iR}B_{jR}C_{kR} \\
  \left\Vert \Gamma^q_{rF} \right\Vert^2
  \quad =& \quad
    \sum_{q,r,F} {\Gamma^\ast}^{rF}_q \Gamma^q_{rF}
\end{align}

\section{Optimized auxiliary field approximation}
\label{sec:OAF}
In this section we discuss how to approximate the Coulomb integrals,
a tensor of fourth order, by a contraction of two considerably smaller tensors
of third order: $V^{pq}_{sr} \approx {\Gamma^\ast}^{pF}_s \Gamma^q_{rF}$,
without actually calculating the entire tensor $V^{pq}_{sr}$.

Given the spin orbitals $\psi_q({\bf x})$ from a Hartree--Fock (HF) or
density functional theory (DFT) calculation, the (nonantisymmetrized)
Coulomb integrals are defined by
\begin{equation}
  V^{pq}_{sr} = \iint{\rm d}{\bf x}\,{\rm d}{\bf x'}\,
  {\psi^\ast}^p({\bf x}) {\psi^\ast}^q({\bf x'})
  \frac1{\left|{\bf r}-{\bf r'}\right|}
  \psi_r({\bf x'}) \psi_s({\bf x})\,,
\end{equation}
with ${\bf x}=({\sigma,\bf r})$ and
$\int{\rm d}{\bf x} = \sum_\sigma\int{\rm d}{\bf r}$.
Owing to the translational invariance of the Coulomb kernel,
we can separate the Coulomb integrals as follows
\begin{equation}
  V^{pq}_{sr} =
  \int\frac{{\rm d}{\bf G}}{(2\pi)^3}\,
  {\Gamma^\ast}^p_s({\bf G}) \Gamma^q_r({\bf G})\,,
  \label{eqn:GammaGammaIntegal}
\end{equation}
where the \emph{Coulomb vertex} $\Gamma^q_r({\bf G})$ is given by
\begin{equation}
  \label{eqn:CoulombVertex}
  \Gamma^q_r({\bf G}) :=
  \sqrt{\frac{4\pi}{{\bf G}^2}}
  \int{\rm d}{\bf x}\,
  {\rm e}^{-{\rm i}{\bf G}\cdot{\bf r}}\,
  {\psi^\ast}^q({\bf x}) \psi_r({\bf x})\,.
\end{equation}
We let its discretization be
$\tilde\Gamma^q_{rG} = \sqrt{w_G}\,\Gamma^q_r({\bf G}_G)$ with the
momentum grid points ${\bf G}_G$ and the numerical integration
weights $w_G$ such that
\begin{equation}
  \int\frac{{\rm d}{\bf G}}{(2\pi)^3}\,
    {\Gamma^\ast}^p_s({\bf G}) \Gamma^q_r({\bf G})
  \approx
  \sum_{G=1}^{N_G} \tilde\Gamma^\ast{}^{pG}_s {\tilde\Gamma}^q_{rG}\,.
  \label{eqn:GammaGammaSum}
\end{equation}
The Coulomb vertex can be computed from the spin orbitals in
$\mathcal O(N_p^2N_G\log N_r)$ time employing a Fast Fourier Transform (FFT),
where $N_p$, $N_G$ and $N_r$ denote the number of spin orbitals, momentum
grid points and real space grid points, respectively.
We note that the orbital overlap charge density
${\psi^\ast}^q({\bf x})\psi_r({\bf x})$
is approximated in the projector augmented wave method using Eq.~(2.87) of
Ref.~\onlinecite{harl_linear_2008}
as implemented in the Vienna \emph{ab initio} simulation package
(\texttt{VASP})~\cite{blochl_projector_1994,kresse_norm-conserving_1994,kresse_efficient_1996}.

In general, any two body operator can be split into a product of two
single body operators coupled by an auxiliary
field~\cite{blankenbecler_monte_1981}.
In the case of the Coulomb interaction the auxiliary field has
only one variable due to translational invariance,
which is the momentum $\bf G$ mediated by the interaction.
Although $\tilde\Gamma^q_{rG}$ is in practice already a third order tensor,
the number of momentum grid points $N_G$ of the HF or DFT calculation is
usually too large to continue with the tensor rank decomposition
of the Coulomb vertex directly.
A large set of momenta will have a small but nonnegligible contribution
to the correlation energy and we seek a more compact set of
auxiliary field variables with fewer relevant elements.

Let
\begin{equation}
  \tilde\Gamma^I_G = U^F_G \Sigma^J_F {W^\ast}^I_J\,,
  \hfill
  \diagramBox{
    \begin{fmffile}{FullGammaMatrix}
    \begin{fmfgraph*}(30,20)
      \fmfstraight
      \fmfleft{G}
      \fmfright{I}
      \fmf{plain}{G,Gamma}
      \fmf{dbl_plain}{Gamma,I}
      \fmfv{d.sh=square,d.f=empty,d.si=16,l=$\tilde\Gamma$,l.d=0.}{Gamma}
      \fmfv{label=$G$,label.angle=105}{G}
      \fmfv{label=$I$,label.angle=75}{I}
    \end{fmfgraph*}
    \end{fmffile}
  }
  \hspace*{-1ex}
  =
  \hspace*{-1ex}
  \diagramBox{
    \begin{fmffile}{GammaSVD}
    \begin{fmfgraph*}(80,20)
      \fmfstraight
      \fmfleft{G}
      \fmfright{I}
      \fmf{plain,tension=0.85}{G,U}
      \fmf{plain}{U,F,S}
      \fmf{dbl_plain}{S,J,V}
      \fmf{dbl_plain,tension=0.85}{V,I}
      \fmfv{d.sh=square,d.f=empty,d.si=16,l=$U$,l.d=0.}{U}
      \fmfv{d.sh=square,d.f=empty,d.si=16,l=$\Sigma$,l.d=0.}{S}
      \fmfv{d.sh=square,d.f=empty,d.si=16,l=$W^\ast$,l.d=0.}{V}
      \fmfv{label=$G$,label.angle=105}{G}
      \fmfv{label=$I$,label.angle=75}{I}
    \end{fmfgraph*}
    \end{fmffile}
  }
  \label{eqn:GammaSVD}
\end{equation}
be a singular value decomposition of the Coulomb vertex $\tilde\Gamma^I_G$,
written as a matrix with the compound index $I=(q,r)$, where
the singular values in $\Sigma$ are sorted in descending order.
Eq.~(\ref{eqn:GammaSVD}) is also shown in form of a wiring diagram
of the involved tensor contractions on the right.
Taking only the largest $N_F<N_G$ singular values of the unapproximated
Coulomb vertex $\tilde\Gamma$ into account
we can define the \emph{optimized auxiliary field (OAF)
Coulomb vertex}
\begin{equation}
  \Gamma^I_F := \Sigma^J_F {W^\ast}^I_J = {U^\ast}^G_F \tilde\Gamma^I_G\,.\hfill
  \diagramBox{
    \begin{fmffile}{GammaMatrix}
    \begin{fmfgraph*}(30,20)
      \fmfstraight
      \fmfleft{F}
      \fmfright{I}
      \fmf{plain}{F,Gamma}
      \fmf{dbl_plain}{Gamma,I}
      \fmfv{d.sh=square,d.f=empty,d.si=16,l=$\Gamma$,l.d=0.}{Gamma}
      \fmfv{label=$F$,label.angle=105}{F}
      \fmfv{label=$I$,label.angle=75}{I}
    \end{fmfgraph*}
    \end{fmffile}
  }
  \hspace*{-0.5ex}
  :=
  \hspace*{-1ex}
  \diagramBox{
    \begin{fmffile}{OptimizedGamma}
    \begin{fmfgraph*}(55,20)
      \fmfstraight
      \fmfleft{F}
      \fmfright{I}
      \fmf{plain,tension=0.85}{F,UT}
      \fmf{plain}{UT,G,Gamma}
      \fmf{dbl_plain,tension=0.85}{Gamma,I}
      \fmfv{d.sh=square,d.f=empty,d.si=16,l=$U^\ast$,l.d=0.}{UT}
      \fmfv{d.sh=square,d.f=empty,d.si=16,l=$\tilde\Gamma$,l.d=0.}{Gamma}
      \fmfv{label=$F$,label.angle=105}{F}
      \fmfv{label=$I$,label.angle=75}{I}
    \end{fmfgraph*}
    \end{fmffile}
  }
  \label{eqn:AofTransform}
\end{equation}
Note that we write
$\Gamma$ without a tilde for the approximated Coulomb vertex, in contrast
to usual convention, simply because we will not use the unapproximated
vertex in any subsequent step.

We are only interested in the left singular vectors $U^F_G$
associated to the largest singular values so we contract
Eq.~(\ref{eqn:GammaSVD}) from the right with $\tilde\Gamma^\ast{}^{G'}_I$
\begin{equation}
  \tilde\Gamma^I_G \tilde\Gamma^\ast{}^{G'}_I =
    U^F_G {\Sigma^2}_F^F {U^\ast}^{G'}_F =: E^{G'}_G,
\end{equation}
transforming a singular value problem of a large $N_G\times N_p^2$ matrix
into an eigenvalue problem of a comparatively small $N_G\times N_G$ hermitian
matrix. The eigenvalues of $E$ are the squares of the singular values of
$\tilde\Gamma^I_G$, and the left eigenvectors of $E$ associated to the largest
eigenvalues are also the left singular vectors of $\tilde\Gamma$ we
need in order to transform the Coulomb vertex $\tilde\Gamma^I_G$ into the
optimized auxiliary field Coulomb vertex $\Gamma^I_F$ according to
Eq.~(\ref{eqn:AofTransform}). Note that this approach becomes numerically
problematic for very small singular values since one only has access to their
squares. However, we find that all $N_F$ largest singular values needed
for an accurate approximation of the Coulomb vertex are sufficiently large.

Inserting the singular value decomposition of the Coulomb vertex
with sorted singular values from Eq.~(\ref{eqn:GammaSVD})
into the discretized definition of
the Coulomb integrals given in Eq.~(\ref{eqn:GammaGammaSum}) yields a
singular value decomposition of the Coulomb integrals, also with
sorted singular values:
\begin{equation}
  V^I_J = W^K_J {\Sigma^2}^K_K {W^\ast}^I_K\,, \hfill
  \diagramBox{
    \begin{fmffile}{FullV}
    \begin{fmfgraph*}(30,20)
      \fmfstraight
      \fmfleft{J}
      \fmfright{I}
      \fmf{dbl_plain}{J,V,I}
      \fmfv{d.sh=square,d.f=empty,d.si=16,l=$V$,l.d=0.}{V}
      \fmfv{label=$J$,label.angle=105}{J}
      \fmfv{label=$I$,label.angle=75}{I}
    \end{fmfgraph*}
    \end{fmffile}
  }
  \hspace*{-1ex}
  =
  \hspace*{-1ex}
  \diagramBox{
    \begin{fmffile}{VSVD}
    \begin{fmfgraph*}(80,20)
      \fmfstraight
      \fmfleft{J}
      \fmfright{I}
      \fmf{dbl_plain,tension=0.85}{J,W}
      \fmf{dbl_plain}{W,K,Sigma,L,WT}
      \fmf{dbl_plain,tension=0.85}{WT,I}
      \fmfv{d.sh=square,d.f=empty,d.si=16,l=$W$,l.d=0.}{W}
      \fmfv{d.sh=square,d.f=empty,d.si=16,l=$\Sigma^2$,l.d=0.}{Sigma}
      \fmfv{d.sh=square,d.f=empty,d.si=16,l=$W^\ast$,l.d=0.}{WT}
      \fmfv{label=$J$,label.angle=105}{J}
      \fmfv{label=$I$,label.angle=75}{I}
    \end{fmfgraph*}
    \end{fmffile}
  }
\end{equation}
where the Coulomb integrals $V^{pq}_{sr}$ are now written in matrix form
$V^I_J$ with $I=(q,r)$ and $J=(s,p)$. Thus, using the optimized auxiliary
field Coulomb vertex $\Gamma^I_F$ instead of the full Coulomb vertex
$\tilde\Gamma^I_G$ best approximates the Coulomb integrals
with respect to the Frobenius norm of the difference:
\begin{equation}
  V^{pq}_{sr} \approx {\Gamma^\ast}^{pF}_s \Gamma^q_{rF}\,. \hfill
  \diagramBox{
    \begin{fmffile}{Vpqrs}
    \begin{fmfgraph*}(50,40)
      \fmfset{arrow_len}{6}
      \fmfstraight
      \fmftop{p,q}
      \fmfbottom{s,r}
      \fmf{fermion,label.side=right,label=$s$}{s,vl}
      \fmf{fermion,label.side=left,label=$r$}{r,vr}
      \fmf{fermion,label.side=right,label=$p$}{vl,p}
      \fmf{fermion,label.side=left,label=$q$}{vr,q}
      \fmf{boson,tension=0.5}{vl,vr}
    \end{fmfgraph*}
    \end{fmffile}
  }  
  \approx
  \hspace*{1ex}
  \diagramBox{
    \begin{fmffile}{GammaGamma}
    \begin{fmfgraph*}(40,40)
      \fmfset{arrow_len}{6}
      \fmfstraight
      \fmfleft{s,p}
      \fmfright{r,q}
      \fmfv{d.sh=square,d.f=empty,d.si=16,l=$\Gamma^\ast$,l.d=0.}{x}
      \fmfv{d.sh=square,d.f=empty,d.si=16,l=$\Gamma$,l.d=0.}{y}
      \fmf{fermion}{s,x}
      \fmf{fermion}{r,y}
      \fmf{fermion}{x,p}
      \fmf{fermion}{y,q}
      \fmffreeze
      \fmf{plain,tension=0.5,label.side=left,label=$F$}{x,y}
      \fmfv{label=$q$,label.angle=-10}{q}
      \fmfv{label=$r$,label.angle=+10}{r}
      \fmfv{label=$p$,label.angle=-170}{p}
      \fmfv{label=$s$,label.angle=170}{s}
    \end{fmfgraph*}
    \end{fmffile}
  }
  \label{eqn:RICoulomb}
\end{equation}

\section{Decomposition of the Coulomb vertex}
\label{sec:TRD}
The form of the Coulomb vertex in real space on the right hand side in
Eq.~(\ref{eqn:CoulombVertex}) suggests
that the optimized Coulomb vertex $\Gamma^q_{rF}$ can be decomposed
in an analogous manner into a product of three tensors of second
order, denoted and depicted as follows
\begin{equation}
\Gamma^q_{rF} \approx \Lambda^R_F{\Pi^\ast}^q_R\Pi^R_r\,.
  \hfill
  \diagramBox{
    \begin{fmffile}{Gamma}
    \begin{fmfgraph*}(20,40)
      \fmfset{arrow_len}{6}
      \fmfstraight
      \fmfright{r,q}
      \fmfleft{x}
      \fmfv{d.sh=square,d.f=empty,d.si=16,l=$\Gamma$,l.d=0.}{y}
      \fmf{fermion}{r,y}
      \fmf{fermion}{y,q}
      \fmf{plain,tension=0.0}{x,y}
      \fmfv{label=$r$,label.angle=0}{r}
      \fmfv{label=$q$,label.angle=0}{q}
      \fmfv{label=$F$,label.angle=-90}{x}
    \end{fmfgraph*}
    \end{fmffile}
  }
  \hspace*{1ex}
  \approx
  \hspace*{-1ex}
  \diagramBox{
    \begin{fmffile}{LambdaPiPiConj}
    \begin{fmfgraph*}(40,60)
      \fmfset{arrow_len}{6}
      \fmfstraight
      \fmfright{r,y,q}
      \fmfleft{x}
      \fmfv{d.sh=square,d.f=empty,d.si=16,l=$\Pi^\ast$,l.d=0.}{qR}
      \fmfv{d.sh=square,d.f=empty,d.si=16,l=$\Pi$,l.d=0.}{rR}
      \fmfv{d.sh=square,d.f=empty,d.si=16,l=$\Lambda$,l.d=0.}{GR}
      \fmf{fermion}{r,rR}
      \fmf{fermion}{qR,q}
      \fmf{plain}{x,GR}
      \fmf{plain}{rR,y}
      \fmf{plain}{qR,y}
      \fmf{plain}{GR,y}
      \fmfv{
        decor.shape=circle,decor.size=4,label.dist=4,label.angle=0,label=$R$
      }{y}
      \fmfv{label=$r$,label.angle=0}{r}
      \fmfv{label=$q$,label.angle=0}{q}
      \fmfv{label=$F$,label.angle=-90}{x}
    \end{fmfgraph*}
    \end{fmffile}
  }
  \label{eqn:GammaConjFactorization}
\end{equation}
We let $N_R$ denote the number of \emph{vertex indices} $R$ and refer to it as
the \emph{rank} of the Coulomb vertex for a given quality of the approximation.
We call the matrices $\Pi_r^R$ and $\Lambda^R_F$ \emph{factor orbitals} and
\emph{Coulomb factors} of the Coulomb vertex, respectively.

The decomposition is invariant under scaling of the Coulomb factors
$\Lambda^R_F$ with any real scalar $a_R>0$ while scaling the
factor orbitals $\Pi_r^R$ with a complex scalar $c_R$ with $|c_R|=1/\sqrt{a_R}$
for each value of $R$.
One can also choose an alternative ansatz to
Eq.~(\ref{eqn:GammaConjFactorization})
for approximating the Coulomb vertex $\Gamma^q_{rF}$
which does not involve the conjugation of the factor orbitals on the
outgoing index $q$. This ansatz reads
\begin{equation}
\Gamma^q_{rF} \approx \Lambda^R_F\Pi^R_q\Pi^R_r\,.
  \hfill
  \diagramBox{
    \begin{fmffile}{Gamma}
    \begin{fmfgraph*}(20,40)
      \fmfset{arrow_len}{6}
      \fmfstraight
      \fmfright{r,q}
      \fmfleft{x}
      \fmfv{d.sh=square,d.f=empty,d.si=16,l=$\Gamma$,l.d=0.}{y}
      \fmf{fermion}{r,y}
      \fmf{fermion}{y,q}
      \fmf{plain,tension=0.0}{x,y}
      \fmfv{label=$r$,label.angle=0}{r}
      \fmfv{label=$q$,label.angle=0}{q}
      \fmfv{label=$F$,label.angle=-90}{x}
    \end{fmfgraph*}
    \end{fmffile}
  }
  \hspace*{1ex}
  \approx
  \hspace*{-1ex}
  \diagramBox{
    \begin{fmffile}{LambdaPiPi}
    \begin{fmfgraph*}(40,60)
      \fmfset{arrow_len}{6}
      \fmfstraight
      \fmfright{r,y,q}
      \fmfleft{x}
      \fmfv{d.sh=square,d.f=empty,d.si=16,l=$\Pi$,l.d=0.}{qR}
      \fmfv{d.sh=square,d.f=empty,d.si=16,l=$\Pi$,l.d=0.}{rR}
      \fmfv{d.sh=square,d.f=empty,d.si=16,l=$\Lambda$,l.d=0.}{GR}
      \fmf{fermion}{r,rR}
      \fmf{fermion}{qR,q}
      \fmf{plain}{x,GR}
      \fmf{plain}{rR,y}
      \fmf{plain}{qR,y}
      \fmf{plain}{GR,y}
      \fmfv{
        decor.shape=circle,decor.size=4,label.dist=4,label.angle=0,label=$R$
      }{y}
      \fmfv{label=$r$,label.angle=0}{r}
      \fmfv{label=$q$,label.angle=0}{q}
      \fmfv{label=$F$,label.angle=-90}{x}
    \end{fmfgraph*}
    \end{fmffile}
  }
  \label{eqn:GammaFactorization}
\end{equation}
The above decomposition is invariant under scaling of the Coulomb factors
$\Lambda^R_F$ with any complex scalar $c_R\neq0$ while scaling the
factor orbitals $\Pi_r^R$ with $\pm1/\sqrt{c_R}$ for each value of $R$.
This is in contrast to the symmetries of the ansatz according to
Eq.~(\ref{eqn:GammaConjFactorization}).
In the Alternating Least Square (ALS) approximation scheme, which we employ for
fitting the factor orbitals $\Pi$ and the Coulomb factors $\Lambda$,
it is preferable
to have the symmetries of Eq.~(\ref{eqn:GammaFactorization}) since
fixing one of the two factors removes all continuous symmetries from the other.
This accelerates convergence and allows a smaller rank $N_R$ in practice.
A downside of this ansatz is that one loses the simple notion of
particle and hole propagators as given in
Eq.~(\ref{eqn:ParticlePropagator}) and Eq.~(\ref{eqn:HolePropagator}),
respectively. The propagators can, for instance, be employed to calculate
second order M\o ller--Plesset theory (MP2) correlation energies in
$\mathcal O(N^4)$, as outlined in Subsection \ref{ssc:Propagators}.
Unless otherwise stated we use the ansatz of
Eq.~(\ref{eqn:GammaFactorization}) for the applications presented
in Section \ref{sec:Results} but we
will continue to discuss the more widely applicable ansatz
of Eq.~(\ref{eqn:GammaConjFactorization}).

If one chooses $N_R=N_r$, where $N_r$ is the number of real space grid points,
the validity of the ansatz follows directly from
Eq.~(\ref{eqn:CoulombVertex}). 
We want to investigate how low $N_R$ can be chosen compared to $N_r$
for a sufficiently
faithful decomposition having an error below 1\% in the energies
calculated from the factor matrices. For reference we use the error
in the MP2 energy assuming that other terms, occurring for instance in the
coupled cluster amplitude equations, exhibit a similar behavior.
This accuracy is assumed sufficient since in practice only those terms
will be calculated from the factor matrices that
pose either computational or memory bottlenecks.
Furthermore, we want to show that the ratio $N_R/N_r$ for a sufficiently
faithful decomposition is independent of the system size if the system is not
too small.

\subsection{Canonical polyadic decomposition algorithms}
A factorization of a tensor according to the ansatz of
Eq.~(\ref{eqn:GammaConjFactorization}) or (\ref{eqn:GammaFactorization})
is referred to as canonical polyadic decomposition~\cite{hitchcock_sum_1927}
(CPD).
For a given rank $N_R$, the factor orbitals $\Pi_r^R$ and the
Coulomb factors $\Lambda^R_F$ can be fit by minimizing the square of the
Frobenius norm of the difference 
\begin{equation}
  \label{eqn:Frobenius}
  (\Lambda, \Pi) = \underset{\Lambda,\,\Pi}{\rm argmin}
  \left\Vert  \Lambda^R_F{\Pi^\ast}^q_R\Pi_r^R - \Gamma^q_{rF} \right\Vert^2.
\end{equation}
The above optimization problem is high dimensional and nonquadratic.
Conjugate gradient algorithms or other local algorithms may require thousands of steps until
sufficiently converged. 
Global optimization algorithms try to tackle the problem by keeping a subset of the
variables fixed and optimizing only the remaining variables.
In the case of the alternating least squares~\cite{kolda_tensor_2009}
(ALS) algorithm the optimization is done in turn over each matrix, while in the
case of the cyclic coordinate descent~\cite{karlsson_parallel_2015} (CCD)
algorithm the optimization is done in turn over each value of the index $R$.
We have studied the performance of a regularized version of the ALS here.

\subsection{Alternating least squares}
In the case of three distinct factors $T_{ijk}\approx A_{iR}B_{jR}C_{kR}$
two of them can be regarded fixed leaving a least squares problem for
finding the optimal third factor. Each matrix is optimized in alternating order
leading to the \emph{alternating least squares} (ALS) algorithm
\begin{align}
  \label{eqn:ALS1}
  A^{(n+1)} &:= \underset{A}{{\rm argmin}}\left\Vert A_{iR}B^{(n)}_{jR}C^{(n)}_{kR} - T_{ijk} \right\Vert^2,\\
  B^{(n+1)} &:= \underset{B}{{\rm argmin}}\left\Vert A^{(n+1)}_{iR}B_{jR}C^{(n)}_{kR} - T_{ijk} \right\Vert^2,\\
  C^{(n+1)} &:= \underset{C}{{\rm argmin}}\left\Vert A^{(n+1)}_{iR}B^{(n+1)}_{jR}C_{kR} - T_{ijk} \right\Vert^2,
\end{align}
which has to be solved iteratively until convergence, starting with random matrices $A^{(0)},B^{(0)}$ and $C^{(0)}$.

Each least squares problem has a unique solution, which can be written
explicitly. For Eq.~(\ref{eqn:ALS1}) it is for instance given by
\begin{equation}
  \label{eqn:ALS}
  A^{(n+1)}_{iR} = T_{ijk} {B^\ast}^{jS} {C^\ast}^{kS} G^+_{SR}\,,
\end{equation}
omitting the iteration specification on $B$ and $C$ for brevity.
$G^+$ denotes the Moore--Penrose
pseudoinverse~\cite{dresden_fourteenth_1920,penrose_generalized_1955}
of the Gramian matrix $G$. For Eq.~(\ref{eqn:ALS1}) the Gramian matrix
is given by
\begin{equation}
  \label{eqn:Gramian}
  G_{RS} = {B^\ast}^{jS} {C^\ast}^{kS} B_{jR} C_{kR}\,.
\end{equation}
The expressions for the other matrices can be written in an analogous manner.
When applying the ALS algorithm to decompose the Coulomb vertex
the computationally most demanding steps are
the calculation of the pseudo inverse
$G^+$ scaling as $\mathcal O(N_R^3)$, as well as the contraction
of $T_{ijk}$ with either factor ${B^\ast}^{jS}$ or ${C^\ast}^{kS}$
in Eq.~(\ref{eqn:ALS}), depending
on which is larger, scaling as $\mathcal O(N_p^2 N_F N_R)$.

\subsection{Regularized alternating least squares}
Although the ALS algorithm guarantees an improvement of the fit quality
in each iteration the convergence can be very slow, especially when
there are multiple minima for a factor $A$ in different regions having
all similar minimal values. In that case the best choice for $A$ may vary
strongly from iteration to iteration since updating the other factors $B$
and $C$ can change the order of the minima. This behavior is referred to
as swamping~\cite{kindermann_analysis_2011} and it takes many iterations
before the ALS algorithm converges to one region for each factor that
globally minimizes the fit quality.
Introducing a penalty on the distance to the previous iteration limits
swamping and leads to the regularized ALS~\cite{cichocki_regularized_2007} (RALS)
algorithm:
\begin{align}
  \nonumber
  A^{(n+1)} :=
    \underset{A}{{\rm argmin}}\Bigg(&
      \left\Vert A_{iR}B^{(n)}_{jR}C^{(n)}_{kR} - T_{ijk} \right\Vert^2 \\
    \label{eqn:RALS1}
    & +\lambda^{(n)}_A\left\Vert A_{iR}-A^{(n)}_{iR}\right\Vert^2\Bigg),
\end{align}
\begin{align}
  \nonumber
  B^{(n+1)} :=
    \underset{B}{{\rm argmin}}\Bigg(&
      \left\Vert A^{(n+1)}_{iR}B_{jR}C^{(n)}_{kR} - T_{ijk} \right\Vert^2 \\
    & +\lambda^{(n)}_B\left\Vert B_{jR}-B^{(n)}_{jR}\right\Vert^2\Bigg),
\end{align}
\begin{align}
  \nonumber
  C^{(n+1)} :=
    \underset{C}{{\rm argmin}}\Bigg(&
      \left\Vert A^{(n+1)}_{iR}B^{(n+1)}_{jR}C_{kR} - T_{ijk} \right\Vert^2 \\
    & +\lambda^{(n)}_C\left\Vert C_{kR}-C^{(n)}_{kR}\right\Vert^2\Bigg).
\end{align}
The solution of each regularized least squares problem can again be given
explicitly, here for instance for
Eq.~(\ref{eqn:RALS1}), and again omitting the iteration specification on
$B$ and $C$
\begin{equation}
  \label{eqn:RALS}
  A^{(n+1)}_{iR} = \left(T_{ijk} {B^\ast}^{jS} {C^\ast}^{kS} +
    \lambda^{(n)}_A A^{(n)}_{iS}\right) G^+_{SR}\,.
\end{equation}
In the regularized case the Gramian $G$ depends on the regularization parameter
$\lambda_A^{(n)}$
\begin{equation}
  \label{eqn:GramianRals}
  G_{RS} = {B^\ast}^{jS} {C^\ast}^{kS} B_{jR} C_{kR} +
    \lambda_A^{(n)}\delta_{RS}\,,
\end{equation}
where $\delta_{RS}$ denotes the Kronecker delta.

The regularization parameter $\lambda_A^{(n)}$ for finding $A^{(n+1)}$
in the $n$th iteration still remains to be determined.
Too low values allow swamping to occur
while too large values unnecessarily slow down the convergence.
To estimate an efficient regularization parameter
we assume that the fit quality $\Vert A B^{(n)}C^{(n)} - T\Vert^2$
in the term to minimize in Eq.~(\ref{eqn:RALS1})
varies little from one iteration to the next. We also assume this for the
local change of the fit quality with respect to each value in $A$.
This allows us to relate the minimized term
in the previous step $n-1$ to the minimized term in the step $n$
for which we want to determine the regularization parameter:
\begin{equation}
  \lambda^{(n-1)} \Vert A^{(n)} - A^{(n-1)} \Vert^2
  \approx
  \lambda^{(n)} \Vert A^{(n+1)} - A^{(n)} \Vert^2.
\end{equation}
If $A^{(n)}$ and $A^{(n+1)}$ have similar norm we can also relate their
relative step sizes $s_A^{(n)}$ and $s_A^{(n+1)}$, by
$
\lambda^{(n-1)} {s^2_A}^{(n)}
  \approx
\lambda^{(n)} {s^2_A}^{(n+1)},
$
where the relative step size in the $n$th iteration is given by
\begin{equation}
  s_A^{(n)} :=
    \Vert A^{(n)} - A^{(n-1)} \Vert /
    \Vert A^{(n)} \Vert.
\end{equation}
We want the relative iteration step size $s_A^{(n+1)}$
of the next iteration to be approximately as large as a chosen maximum
value $s_0$, which we refer to as \emph{swamping threshold}.
From that we define the estimated
regularization parameter for the $n$th iteration
\begin{equation}
  {\hat\lambda}_A^{(n)} := \lambda_A^{(n-1)} {s^2_A}^{(n)}/s^2_0.
\end{equation}
Using the above estimate directly results in a regularization which we find
alternately too strong and too weak.
To ameliorate this we introduce a mixing of the estimated regularization
parameter $\hat\lambda_A^{(n)}$ for the $n$th iteration, as above, with
the regularization parameter $\lambda_A^{(n-1)}$ of the previous iteration
to obtain the regularization parameter $\lambda_A^{(n)}$ employed
for the $n$th iteration in the RALS:
\begin{align}
  \lambda_A^{(0)} &:= 1 \\
  \lambda_A^{(n)} &:=
    \alpha{\hat\lambda}_A^{(n)} + (1-\alpha)\lambda_A^{(n-1)}.
  \label{eqn:RegularizationParameter}
\end{align}
Regarding the choice of the swamping threshold $s_0$ and the mixing factor
$\alpha$ we find that $s_0=1.0$ and $\alpha=0.8$
offers a good compromise allowing quick convergence
while still preventing swamping for the systems studied so far.

\subsection{Quadratically occurring factors}
In the case of the Coulomb vertex the factor orbitals $\Pi_r^R$ occur
quadratically.
For finding the next estimate $\Pi^{(n+1)}$ in the alternating least squares
algorithm we use an iterative algorithm
similar to the Babylonian square root algorithm. Each subiteration is
given by
\begin{align}
  \nonumber
  \Pi^{(n+1,m+1)} := (&1-\beta)\,\Pi^{(n+1,m)}\\
  \nonumber
    +\beta\,\underset{\Pi}{{\rm argmin}}\Bigg(&
      \left\Vert {\Lambda^R_F}^{(n+1)}{{\Pi^\ast}^q_R}^{(n+1,m)}\Pi_r^R -
        \Gamma^q_{rF} \right\Vert^2 \\
       &+ \lambda_\Pi^{(n+1,m)}\left\Vert \Pi_r^R - {\Pi^R_r}^{(n+1,m)} \right\Vert
    \Bigg)
\end{align}
with $\Pi^{(n+1,0)}:=\Pi^{(n)}$ and the mixing factor $0<\beta<1$.
Note that $\Pi^\ast$ is a fixed parameter rather than a fitted one
and that the regularization parameter $\lambda_\Pi^{(n+1,m)}$ needs
to be determined for the $m$th subiteration similar to
Eq.~(\ref{eqn:RegularizationParameter}), however with
$\lambda_\Pi^{(n+1,0)}=\lambda_\Pi^{(n)}$.
The above iteration converges towards a solution
of the quadratic problem. We use $\Pi^{(n+1,M)}$ and $\lambda_\Pi^{(n+1,M)}$
for the next estimate of $\Pi^{(n+1)}$ and $\lambda_\Pi^{(n+1)}$ in the RALS
algorithm, respectively. The number of subiterations
$M$ needs to be sufficiently large, such
that $\Pi^{(n+1)}$ is at least an improved solution of the entire fit problem
compared to $\Pi^{(n)}$. A large number $M$ of subiterations gives
an estimate $\Pi^{(n+1)}$ that is close to the optimal choice of $\Pi$
for a given $\Lambda^{(n+1)}$.
However, the cost
of each subiteration are similar to the cost of the fit of $\Lambda$
in the RALS algorithm
and as a good choice for minimizing the overall computational cost
we find $\beta=0.8$ and $M\geq2$, but only as large
such that the solution is an improvement.
We point out that there are alternative methods for solving the quadratically
occurring factors~\cite{li_convergence_2011}.

\section{Application of the low rank factorization}
\label{sec:Application}
The algorithms described so far yield an approximate factorization of the
Coulomb integrals of the form
\begin{equation}
  V^{pq}_{sr} \approx
    {\Pi^\ast}^p_R {\Pi^\ast}^q_S {\Lambda^\ast}^F_R
    \Lambda^S_F \Pi^S_r \Pi^R_s,
\end{equation}
where the factors $\Pi$ and $\Lambda$ are $N_p\times N_R$ and $N_F\times N_R$
matrices, respectively. We find that the rank of the decomposition $N_R$ is
about an order of magnitude lower than the number of real space grid points
of the original factors of the Coulomb integrals, being the
orbitals $\psi_q({\bf x})$.
In Section \ref{sec:Results} we study the convergence of the approximation
in detail.
In this section we discuss how this factorization can be applied to
lower the scaling of the computational cost of wave function based methods such as
second order M\o ller--Plesset (MP2) theory or coupled cluster theory.

\subsection{MP2 from imaginary time propagators}
\label{ssc:Propagators}
The factorization of the Coulomb integrals permits
evaluating the terms in the perturbation expansion by summing
over all vertex indices $R,S,T,\ldots$ occurring in the term's diagram,
contracting propagator matrices for each particle, hole and Coulomb line.
For instance, the exchange term of second order M\o ller--Plesset (MP2)
theory can be evaluated as follows:
\begin{align}
  \nonumber
  \diagramBox{
    \begin{fmffile}{MP2xGreens}
      \begin{fmfgraph*}(40,40)
        \fmfset{arrow_len}{6pt}
        \fmfleft{v00,v01}
        \fmfright{v10,v11}
        \fmf{boson}{v00,v10}
        \fmf{boson}{v01,v11}
        \fmf{fermion,left=0.25}{v00,v01}
        \fmf{fermion,right=0.25}{v10,v11}
        \fmf{fermion}{v01,m} \fmf{plain}{m,v10}
        \fmf{fermion}{v11,m} \fmf{plain}{m,v00}
        \fmfv{label=$R$,label.dist=4,label.angle=170}{v01}
        \fmfv{label=$S$,label.dist=4,label.angle=10}{v11}
        \fmfv{label=$T$,label.dist=4,label.angle=-170}{v00}
        \fmfv{label=$U$,label.dist=4,label.angle=-10}{v10}
      \end{fmfgraph*}
    \end{fmffile}
  }
  =
  -\frac12
  \int_0^\infty{\rm d}\tau
  \sum_{RSTU}
  & G_T^R(\tau) G_R^U(-\tau) V_R^S \\[-2.5ex]
  & G_U^S(\tau) G_S^T(-\tau) V_T^U,
\end{align}
where the imaginary time dependent propagator matrices are given
in terms of the decomposed factor orbitals $\Pi^R_r$ and the Coulomb factors
$\Lambda^R_F$:
\begin{align}
  G^R_S(\tau\geq0) &:=
    +\sum_a{\Pi^\ast}_S^a \Pi_a^R\, {\rm e}^{-(\eps_a-\mu)\tau}
  \label{eqn:ParticlePropagator}
  \\
  G^R_S(\tau<0) &:=
    -\sum_i{\Pi^\ast}_S^i \Pi_i^R\, {\rm e}^{-(\eps_i-\mu)\tau}
  \label{eqn:HolePropagator}
  \\
  V^S_R &:= {\Lambda^\ast}^F_R\Lambda^S_F\,,
  \label{eqn:MP2Propagators}
\end{align}
where $\mu$ is the Fermi level energy at zero temperature.
The imaginary time integration can be done numerically on a small grid,
for instance the minimax grid~\cite{kaltak_low_2014},
which allows an evaluation of the MP2 correlation energy in $\mathcal O(N^4)$
in principle. In practice, however, this approach outperforms
canonical MP2 calculations only for very large systems.
The propagator matrices $G^R_S(\pm\tau)$ are analogous to
the one body particle/hole Green's functions in real space and imaginary time
$G_0({\bf x},{\bf x'};\pm\tau)$.
The time independent matrix propagator $V^S_R$ corresponds to the
Coulomb kernel. The factorization of the one body energies $\varepsilon_{a,i}$
in imaginary
time is equivalent to the Laplace transformed MP2 ansatz of
Alml\"of~\cite{almlof_elimination_1991}.
Note that the imaginary time dependent matrices defined above are
actually not propagators in the sense that
\begin{equation}
  G^S_T(\tau_1) G^R_S(\tau_2) = G^R_T(\tau_1+\tau_2)
\end{equation}
for all $R,T$ and $\tau_1,\tau_2>0$ for particles as well as
$\tau_1,\tau_2\leq0$ for holes.
They can only be used to directly connect vertices of the Coulomb interaction.
If propagators in the above sense are required, one needs to employ a
stricter ansatz for the factorization of the Coulomb vertex, namely
\begin{equation}
  \Gamma^q_{rF} \approx \Lambda^R_F {\Pi^+}^q_R \Pi^R_r\,,
\end{equation}
where $\Pi^+$ denotes the Moore--Penrose pseudo inverse of $\Pi$.
The convergence behavior of this ansatz remains, however, to be studied.

One can also evaluate the perturbation terms stochastically,
directly using the real space Green's functions
$G_0({\bf x},{\bf x'};\pm\tau)$ 
rather than using the low rank propagators $G^R_S(\pm\tau)$.
This has been done for one dimensional
solids~\cite{willow_stochastic_2012,neuhauser_stochastic_2016} and for
three dimensional solids~\cite{schaefer_mp2_2016}.

\subsection{Reduced scaling coupled cluster theory}
The most demanding step in the canonical CCD (DCD) method using a plane wave basis set
is the calculation of the particle/particle ladder contribution $T^{cd}_{ij}V^{ab}_{cd}$
in the amplitude equation, scaling as $\mathcal O(N_o^2N_v^4)$ in time.
Here, the factorized
form of the Coulomb integrals $V^{ab}_{cd}$ can be exploited to break down
the simultaneous contraction over the indices $c$ and $d$ into a sequence
of contractions involving only at most one index, as can be seen
from the wiring diagram of the involved tensors:
\begin{equation}
  \diagramBox{
    \begin{fmffile}{DCDdirect}
    \begin{fmfgraph*}(60,40)
      \fmfset{arrow_len}{6}
      \fmfstraight
      \fmfbottom{v00,v10,a,v20,v30,v40}
      \fmftop{v01,v11,b,v21,v31,v41}
      \fmf{plain,label=$T^{cd}_{ij}$,label.side=right}{v10,v30}
      \fmf{fermion,right=0.12}{v01,v10}
        \fmf{fermion}{v10,R,v11}
      \fmf{fermion,left=0.12}{v41,v30}
        \fmf{fermion}{v30,S,v31}
      \fmf{boson,label=$V^{ab}_{cd}$,label.side=left}{R,S}
      \fmfv{label=$i$,label.angle=90,label.dist=3}{v01}
      \fmfv{label=$a$,label.angle=90,label.dist=3}{v11}
      \fmfv{label=$b$,label.angle=90,label.dist=3}{v31}
      \fmfv{label=$j$,label.angle=90,label.dist=2}{v41}
    \end{fmfgraph*}
    \end{fmffile}
  }
  \approx
  \diagramBox{
    \begin{fmffile}{DCDfactors0}
    \begin{fmfgraph*}(100,80)
      \fmfset{arrow_len}{6}
      \fmfstraight
      \fmfbottom{v00,v10,a,v20,v30,v40}
      \fmftop{v01,v11,b,v21,v31,v41}
      \fmf{plain}{v10,v30}
      \fmf{fermion,tension=2,right=0.15}{v01,v10}
        \fmf{fermion,tension=2,label=$c$,l.dist=2,l.side=right}{v10,cR}
        \fmf{plain,tension=2}{cR,R,Ra}
        \fmf{fermion,tension=2}{Ra,v11}
      \fmf{fermion,tension=2,left=0.15}{v41,v30}
        \fmf{fermion,tension=2}{v30,dS}
        \fmf{plain,tension=2}{dS,S,Sb}
        \fmf{fermion,tension=2}{Sb,v31}
      \fmf{plain,tension=2}{R,RS,S}
      \fmfv{d.sh=square,d.f=empty,d.si=16,l=$\Pi$,l.d=0.}{cR}
      \fmfv{d.sh=square,d.f=empty,d.si=16,l=$\Pi^\ast$,l.d=0.}{Ra}
      \fmfv{d.sh=square,d.f=empty,d.si=16,l=$\Pi$,l.d=0.}{dS}
      \fmfv{d.sh=square,d.f=empty,d.si=16,l=$\Pi^\ast$,l.d=0.}{Sb}
      \fmfv{d.sh=square,d.f=empty,d.si=20,l=$\Lambda^\ast\Lambda$,l.d=0.}{RS}
      \fmfv{label=$i$,label.angle=180,label.dist=4}{v01}
      \fmfv{label=$a$,label.angle=0,label.dist=4}{v11}
      \fmfv{label=$b$,label.angle=180,label.dist=4}{v31}
      \fmfv{label=$j$,label.angle=00,label.dist=4}{v41}
      \fmfv{label=$R$,label.angle=180}{R}
      \fmfv{label=$S$,label.angle=0}{S}
      \fmfv{label=$d$,label.angle=157,label.dist=7}{v30}
    \end{fmfgraph*}
    \end{fmffile}
  }
  \label{eqn:LadderFactorization}
\end{equation}
The most expensive term in this sequence of contractions leads to a scaling of
$\mathcal O(N_o^2N_vN_R^2)$ in time, without exceeding the memory complexity
of the coupled cluster amplitudes.
As will be demonstrated in Section \ref{sec:Results}, we find $N_R$ to be
proportional to the system size $N$, resulting in an $\mathcal O(N^5)$
scaling behavior in time of the particle/particle ladder contribution.
Furthermore, the DCD amplitudes equations can be
solely reformulated in an $\mathcal O(N^5)$ implementation with the use of the
Coulomb vertex and its decomposed approximation, due to the absence of
exchange terms between different clusters. Likewise, the most expensive
term in CCSD (DCSD) amplitude equations includes the singles contribution to
the particle/particle ladder diagram ($T^{cd}_{ij}V^{ab}_{ck}T^{k}_{d}$,
$T^{cd}_{ij}V^{ab}_{ld}T^{l}_{c}$, $T^{cd}_{ij}V^{ab}_{kl}T^{k}_{d}T^{l}_{c}$).
Similarly, these terms can be evaluated
via the factor orbitals and the Coulomb factors in an
$\mathcal O(N^5)$ scaling in time and in the DCSD approximation
no term exceeds this scaling behavior.

\section{Results employing the low rank factorization}
\label{sec:Results}

\subsection{Total energies of the LiH solid}
\label{ssc:LiH}

\begin{figure*}
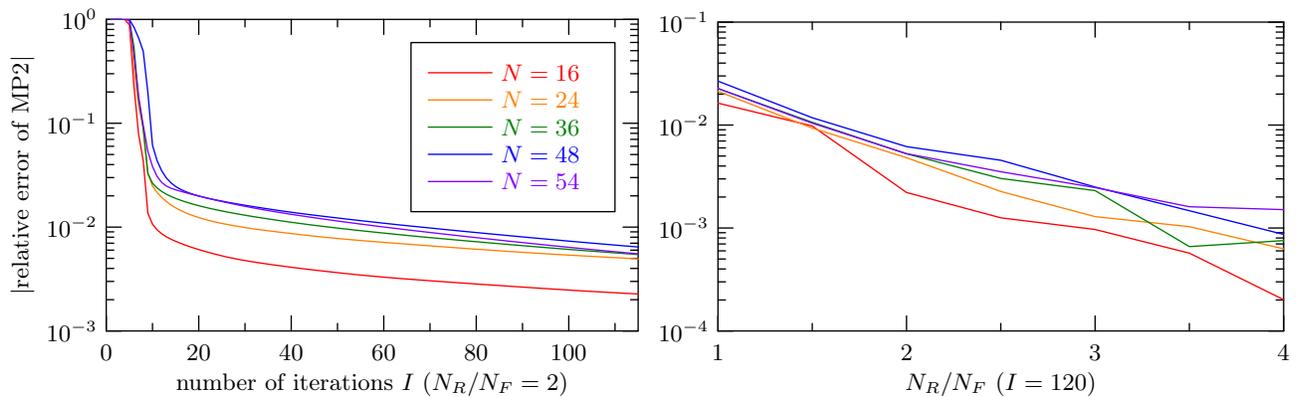

\begin{center}
\begin{asy}
  real[][] it_eps8 = input("it_eps_rank-2_size-8.dat").line().dimension(0,0);
  real[][] it_eps12 = input("it_eps_rank-2_size-12.dat").line().dimension(0,0);
  real[][] it_eps18 = input("it_eps_rank-2_size-18.dat").line().dimension(0,0);
  real[][] it_eps24 = input("it_eps_rank-2_size-24.dat").line().dimension(0,0);
  real[][] it_eps27 = input("it_eps_rank-2_size-27.dat").line().dimension(0,0);

  sizeRatio(width=240);
  scale(Linear, Log);
  plotXY(it_eps8, red, "\small $N=16$");
  plotXY(it_eps12, orange, "\small $N=24$");
  plotXY(it_eps18, deepgreen, "\small $N=36$");
  plotXY(it_eps24, blue, "\small $N=48$");
  plotXY(it_eps27, purple, "\small $N=54$");
  Label l; l.p = fontsize(9);
  axisXY(
    Label("\small number of iterations $I$ ($N_R/N_F=2$)",MidPoint),
    Label("\small$|$relative error of MP2$|$",MidPoint),
    (0,1e-3), (115,1),
    xTicks=LeftTicks(l, Step=20,n=2),
    yTicks=RightTicks(l, N=1,n=10)
  );
  attach(legend(1,6,6,24,vskip=1),(115,0.0),12SW,UnFill);
\end{asy}
\begin{asy}
  sizeRatio(width=240);
  real[][] rank_eps8 = input("rank_eps_size-8.dat").line().dimension(0,0);
  real[][] rank_eps12 = input("rank_eps_size-12.dat").line().dimension(0,0);
  real[][] rank_eps18 = input("rank_eps_size-18.dat").line().dimension(0,0);
  real[][] rank_eps24 = input("rank_eps_size-24.dat").line().dimension(0,0);
  real[][] rank_eps27 = input("rank_eps_size-27.dat").line().dimension(0,0);
  scale(Linear, Log);
  plotXY(rank_eps8, red);
  plotXY(rank_eps12, orange);
  plotXY(rank_eps18, deepgreen);
  plotXY(rank_eps24, blue);
  plotXY(rank_eps27, purple);
  Label l; l.p = fontsize(9);
  axisXY(
    Label("\small$N_R/N_F$ ($I=120$)",MidPoint),
    Label("",MidPoint),
    (1,1e-4), (4,1e-1),
    xTicks=LeftTicks(l, Step=1,n=2),
    yTicks=RightTicks(l, N=1,n=10)
  );
\end{asy}
\end{center}
\vspace*{-2ex}
\caption{
  On the left the convergence of the relative error of the MP2 energy for $N_R=2N_F$ with
  respect to fit iterations for different system sizes is shown.
  The convergence with respect to the rank $N_R$ using
  120 iterations is given on the right.
}
\label{fig:RankConvergence}
\end{figure*}

We first seek to discuss the convergence of the low rank factorization with respect
to the number of iterations, $N_R$ and the system size. To this end we
study different supercell sizes constructed from two atomic LiH crystal unit cells including
2$\times$2$\times$2,
3$\times$2$\times$2,
3$\times$3$\times$2,
4$\times$3$\times$2 and
3$\times$3$\times$3, corresponding to 16, 24, 36, 48 and 54 atoms, respectively.
In this subsection we only employ MP2 theory to investigate the behavior
of the correlation energy calculated from the factorized Coulomb integrals.
This study focuses on the decomposition of the Coulomb vertex, neither
employing the optimized auxiliary field nor the pseudized Gaussian type virtual
orbitals~\cite{booth_plane_2016} technique.
The resulting Coulomb vertices that need to be fit are large, such
that only the particle/hole part $\Gamma^a_{iF}$ is used for this study.
The kinetic energy cutoff defining $N_F$ was set to 200~eV.
The Li 2$s^1$ and H 1$s^1$ states have been treated as valence
states.

Fig.~\ref{fig:RankConvergence} shows the relative error of the MP2 correlation energy retrieved
as a function of the number of iterations.
The relative error is computed from MP2 energies that employ integrals that have been
calculated with and without the low rank tensor factorization.
Fig.~\ref{fig:RankConvergence} reveals that the rate of convergence for the relative error of the MP2 energy
is very similar in all different system sizes. We note that the smallest supercell containing 16 atoms only
exhibits a slightly faster convergence.
From these results we conclude that the required number of iterations in the tensor factorization algorithm is
system size independent for intensive properties. Furthermore we find that 100 iterations are sufficient to achieve
a relative accuracy of 1\% with a rank that corresponds to 2$N_F$.

The right side of Fig.~\ref{fig:RankConvergence} explores the convergence of the relative error in the MP2 correlation
energy error retrieved as a function of $N_R/N_F$. Note that $N_F$ corresponds to the number of plane wave vectors and scales
linearly with respect to system size. This plot shows that the required rank $N_R$ needed to achieve a certain
relative level of accuracy also scales linearly with respect to the system size. Furthermore we find that systematically
improvable exponential convergence can be achieved for this system and property by increasing $N_R/N_F$.

The computational cost for obtaining the TRD of the Coulomb vertex with $N_{\rm o}=27$, $N_{\rm v}=8469$, $N_F=N_G=1830$ and $N_R=1830$
is roughly 1000 CPU hours. Therefore the computational cost of the TRD exceeds the computational cost of a full MP2 calculation
which is roughly 10 CPU hours in the present case despite the fact that the TRD formally scales more favorably with system size.
However, we note that the present TRD algorithm
is sufficiently efficient to reduce the total computational cost of coupled cluster theory calculations
as discussed in the following.

\subsection{Molecular adsorption of water on hexagonal boron nitride}
\label{ssc:Adsorptions}

We now turn to the application of the newly developed methodologies to some
more challenging problems. We calculate the interaction between a water
molecule and a hexagonal boron nitride (\textit{h}BN)
monolayer. We employ periodic coupled cluster doubles (CCD) and
examine to what extent the TRD and the optimized auxiliary field
approximations are accurate and efficient.
We used the structures obtained by Al-Hamdani
\textit{et. al.}~\cite{al-hamdani_communication:_2015},
whereby the molecule is oriented on top
of an N site and the geometry has been optimized using the optB86b-vdW
functional. The water--N distance was set to 3.2~\AA. The \textit{h}BN
monolayer is modeled by 32 atoms in the periodic cell and the distance between two BN sheets
was set to 16~\AA. After checking convergence, we employed a 500~eV kinetic energy
cutoff for the one particle orbitals along with $\Gamma$ point sampling of
the Brillouin zone. 
The B 2$s^22p^1$, N 2$s^22p^3$, O 2$s^22p^4$, and H $1s^1$ states have
been treated as valence states.
Occupied HF states were converged within the full
plane wave basis, whereas the virtual orbitals were constructed using
Dunning's contracted aug-cc-pVDZ (AVDZ) and aug-cc-pVTZ
(AVTZ)~\cite{jr_gaussian_1989,feller_role_1996} pseudized Gaussians in a plane wave
representation, orthogonalized to the occupied
space~\cite{booth_plane_2016}. Pseudized Gaussians
have proven to work reliably and
efficiently for surface studies on periodic systems~\cite{booth_plane_2016}. 
Counterpoise corrections to the basis set superposition error
(BSSE) were included in all correlated calculations.
The adsorption energy is defined as the difference in energy between the
noninteracting fragments and the interacting system
\begin{equation}
  E_{\rm{ads}} =
  E_{\rm{H_2O}} +
  E_{\rm{BN}} -
  E_{\rm{H_2O+BN}}.
\end{equation}
We note that in Ref.~\onlinecite{al-hamdani_communication:_2015} the
adsorption energy has been calculated as the difference between the 
total energy of water and \textit{h}BN at the largest possible
oxygen--surface distance
of 8~\AA~and the total energy of water and \textit{h}BN at the adsorption
oxygen--surface distance.

\begin{figure}
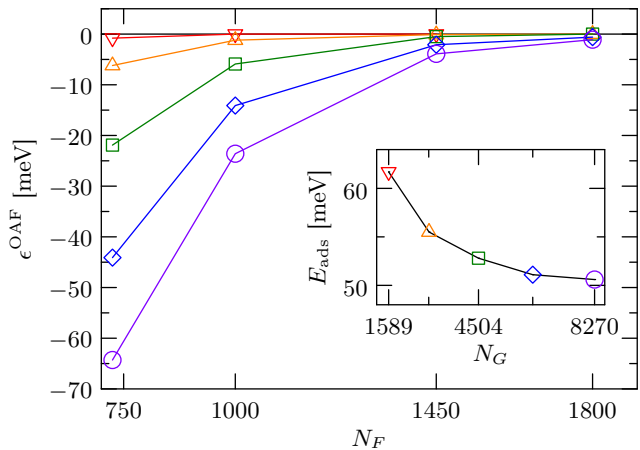

\begin{center}
\begin{asy}
  real[][] ecut100 = input("ecut100.dat").line().dimension(0,0);
  real[][] ecut150 = input("ecut150.dat").line().dimension(0,0);
  real[][] ecut200 = input("ecut200.dat").line().dimension(0,0);
  real[][] ecut250 = input("ecut250.dat").line().dimension(0,0);
  real[][] ecut300 = input("ecut300.dat").line().dimension(0,0);

  sizeRatio(width=240,ratio=sqrt(2));
  path[][] markers = {
    scale(0.8mm) * rotate(180)*scale(sqrt(2))*polygon(3),
    scale(0.8mm) * scale(sqrt(2))*polygon(3),
    scale(0.8mm) * scale(sqrt(2))*polygon(4),
    scale(0.8mm) * rotate(45)*scale(sqrt(2))*polygon(4),
    scale(0.8mm) * scale(sqrt(2))*unitcircle
  };
  pen[] colors = { red, orange, deepgreen, blue, purple };
  draw((700,0)--(1900,0));
  plotXY(ecut100, colors[0], "\small$NG=1589$",marker(markers[0],colors[0]));
  plotXY(ecut150, colors[1], "\small$NG=2899$",marker(markers[1],colors[1]));
  plotXY(ecut200, colors[2], "\small$NG=4504$",marker(markers[2],colors[2]));
  plotXY(ecut250, colors[3], "\small$NG=6264$",marker(markers[3],colors[3]));
  plotXY(ecut300, colors[4], "\small$NG=8270$",marker(markers[4],colors[4]));
  Label l; l.p = fontsize(9);
  axisXY(
    Label("\small$N_F$",MidPoint),
    Label("\small$\epsilon^{\rm OAF}$ [meV]",MidPoint),
    (700,-70), (1900,5),
    xTicks=LeftTicks(l, Ticks=new real[]{750,1000,1450,1800}),
    yTicks=RightTicks(l, Step=10,step=5)
  );

  real[][] ecutgw = input("ecutgw.dat").line().dimension(0,0);
  picture insetPic;
  sizeRatio(insetPic, width=0.5*240, ratio=sqrt(2));
  plotXY(insetPic, ecutgw, black);
  draw(insetPic, (ecutgw[0][0],ecutgw[0][1]), marker(markers[0],colors[0]));
  draw(insetPic, (ecutgw[1][0],ecutgw[1][1]), marker(markers[1],colors[1]));
  draw(insetPic, (ecutgw[2][0],ecutgw[2][1]), marker(markers[2],colors[2]));
  draw(insetPic, (ecutgw[3][0],ecutgw[3][1]), marker(markers[3],colors[3]));
  draw(insetPic, (ecutgw[4][0],ecutgw[4][1]), marker(markers[4],colors[4]));
  Label l; l.p = fontsize(9);
  axisXY(
    insetPic,
    Label("\small$N_G$",MidPoint),
    Label("\small$E_{\rm ads}$ [meV]",MidPoint),
    (1200,48), (8600,64),
    xTicks=LeftTicks(l, Ticks=new real[]{1589,4504,8270},ticks=new real[]{2889,6264}),
    yTicks=RightTicks(l, Step=10,step=5)
  );
  add(insetPic.fit(), (1900,-70), 8*unit(NW));
\end{asy}
\end{center}
\vspace*{-2ex}
\caption{
  Optimized auxiliary field (OAF) approximation error 
  $\epsilon^{\rm OAF}(N_F) = E_{\rm ads}(N_F) - E_{\rm ads}(N_G)$
  of the CCD adsorption
  energy as a function of the number of field variables $N_F$ used to
  approximate the Coulomb vertex. The number of $G$ vectors $N_G$
  of the unapproximated Coulomb vertex ${\tilde\Gamma}^q_{rG}$
  is indicated by the shape of the markers.
  The inset shows which marker corresponds to which $N_G$ and plots
  the convergence of the adsorption energy with respect to
  $N_G$, corresponding to kinetic energy cutoff values
  of 100, 150, 200, 250, and 300~eV, respectively.
}
\label{fig:SVDConvergence}
\end{figure}

Initially, we investigate the convergence of the adsorption energy with
respect to the number of momentum grid points $N_G$, employed to evaluate the
Coulomb vertex ${\tilde\Gamma}^q_{rG}$ according to Eq.~\eqref{eqn:GammaGammaSum}.
The selection of the plane waves vectors $\mathbf{G}$ is determined by a kinetic
energy cutoff $E_{\chi}$ such that
\begin{equation}
  \frac{\hbar^2 \mathbf{G}^2}{2m_e} < E_{\chi}.
  \label{eqn:CutoffEnergy}
\end{equation}
For this purpose we utilize the pseudized
AVDZ basis set for the virtual orbitals.
The current system consists of $N_o=68$ and $N_v=780$ occupied and virtual
orbitals respectively.
Kinetic energy cutoff values from 100 to 300~eV were
employed for the calculation of the adsorption energy. The results are shown
in the inset of
Fig.~\ref{fig:SVDConvergence}. The adsorption energy behaves as
$CE_{\chi}^{-3/2}$~\cite{harl_cohesive_2008,marsman_second-order_2009} up to 200~eV, however, at higher
cutoffs one observes a plateau in the curve as a result of the truncation of
the virtual orbital space via the pseudized Gaussian basis functions.
We conclude that a cutoff energy of 200~eV is sufficient to converge
the adsorption energy to within 2--3~meV.
We then investigate how accurately can the optimized auxiliary field
approximate the Coulomb vertex $\tilde\Gamma^q_{rG}$ for the current
system at the level of CCD theory. First we obtain the optimized Coulomb
vertex $\Gamma^q_{rF} = {U^\ast}_F^G \tilde\Gamma^q_{rG}$, where $U_G^F$
consists of the left singular values of $\tilde\Gamma^q_{rG}$ associated
to the $N_F$ largest singular values, according to Eq.~(\ref{eqn:GammaSVD}).
The optimized Coulomb vertex is expected to be efficient
since most of the space in the simulation cell is vacant, and the plane wave
auxiliary basis contains redundant information. Different number of field
variables $N_F$ were employed to approximate the plane wave vectors $N_G$ for
the various cutoff energies. The behavior of the adsorption energy with respect
to the number of field variables is shown in
Fig.~\ref{fig:SVDConvergence}. The rapid convergence of the energy with
increasing number of field variables owes to the locality of the molecular
orbitals in the supercell. The adsorption energy obtained with a cutoff of
200~eV can be calculated within 0.5~meV accuracy using $N_F=1450$ field
variables to approximate $N_G=4504$ plane wave vectors.

\begin{figure}
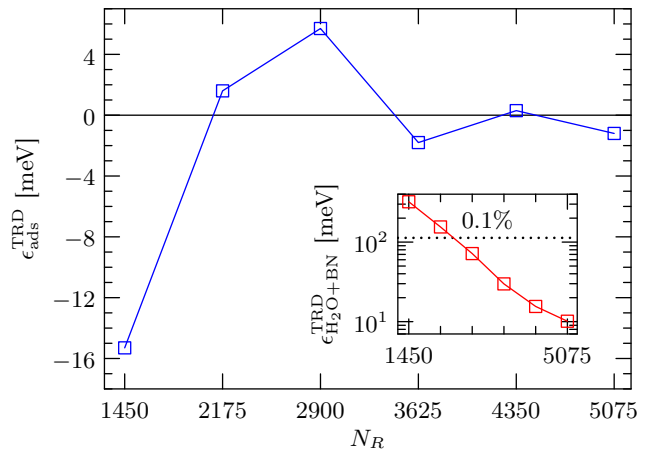

\begin{center}
\begin{asy}
  real[][] trd = input("trd.dat").line().dimension(0,0);

  sizeRatio(width=240,ratio=sqrt(2));
  path[] square = shift(-1,-1)*scale(2)*unitsquare;
  draw((1300,0)--(5200,0));
  plotXY(trd, blue, marker(scale(0.8mm)*square, blue));
  Label l; l.p = fontsize(9);
  axisXY(
    Label("\small$N_R$",MidPoint),
    Label(
      "\small$\epsilon^{\rm TRD}_{\rm ads}$ [meV]",
      MidPoint
    ),
    (1300,-18), (5200,7),
    xTicks=LeftTicks(l, Ticks=new real[]{1450,2175,2900,3625,4350,5075}),
    yTicks=RightTicks(l, Step=4,step=1)
  );

  picture insetPic;
  scale(insetPic, x=Linear, y=Log);
  real[][] ads_trd = input("ads_trd.dat").line().dimension(0,0);
  sizeRatio(insetPic, width=0.45*240,ratio=sqrt(2));
  // draw 0.1
  real permille = 113;
  draw(
    insetPic, Label("\small$0.1\%$",align=N),
    Scale(insetPic,(1200,permille))--Scale(insetPic,(5300,permille)),
    Dotted()
  );
  plotXY(insetPic, ads_trd, red,marker(scale(0.8mm)*square,red));
  axisXY(
    insetPic,
    "",
    Label("\small$\epsilon^{\rm TRD}_{\rm H_2O+BN}$ [meV]",MidPoint),
    (1200,7), (5300,400),
    xTicks=LeftTicks(l, Ticks=new real[]{1450,5075},ticks=new real[]{2175,2900,3625,4350}),
    yTicks=RightTicks(l,N=1,n=10)
  );
  add(insetPic.fit(), (5100,-18), 12NW);
\end{asy}
\end{center}
\vspace*{-2ex}
\caption{
  Low rank approximation error
  $\epsilon^{\rm TRD}_{\rm ads}(N_R)=E_{\rm ads}(N_F,N_R)-E_{\rm ads}(N_F)$
  of the CCD adsorption energy
  as a function of the rank $N_R$ using the OAF approximated Coulomb
  vertex with $N_F=1450$ field variables.
  The inset shows the respective approximation error for surface+molecule
  fragment
  $\epsilon^{\rm TRD}_{\rm H_2O+BN}=E_{\rm H_2O+BN}(N_F,N_R)-E_{\rm H_2O+BN}(N_F)$
  only, revealing an error cancellation of one to two orders of magnitude
  in the adsorption energy.
}
\label{fig:trd}
\end{figure}

Since we can conclude that the adsorption energy can be computed within
approximately~3~meV using a
cutoff energy of~200~eV for the auxiliary plane wave basis and $N_F=1450$ field
variables to construct the optimized Coulomb vertex, we chose these settings
to assess the accuracy of the low rank factorization of the Coulomb
integrals. We used different number of vertex indices $R$ to compute
the factor orbitals and Coulomb factors of the Coulomb vertex,
following Eq.~(\ref{eqn:GammaConjFactorization}).
We approximate the particle/particle ladder contribution $T^{cd}_{ij}V^{ab}_{cd}$
in the amplitude equation of CCD via the factor orbitals and the
Coulomb factors as shown in Eq.~(\ref{eqn:LadderFactorization}). The
adsorption energy versus the number of the decomposed vertex indices $N_R$ is
shown in Fig.~\ref{fig:trd}. The energy does not converge monotonically as in
the case of the optimized auxiliary field approximation. 
This is due to the nonlinear nature of the canonical polyadic decomposition
of the Coulomb vertex and the random initial choice for its factors.
Nevertheless we observe a converged behavior with increasing
$N_R$. Furthermore, a value of $N_R=4350=3N_F$ is sufficient to yield an
adsorption energy within 1~meV accuracy. This suggests that the TRD of the
Coulomb vertex is a controllable approximation that can yield increasingly
accurate results with increasing decomposition rank $N_R$.
In order to further validate the accuracy of the TRD method we show
the convergence of the absolute energy of the interacting system with
respect to the decomposition rank $N_R$ in the inset of
Fig.~\ref{fig:trd}. We observe an exponential convergence of
the total energy. An accuracy better than 0.1\% is achieved already
with $N_R=2N_F$. Nevertheless we stress that the corresponding
accuracy in the adsorption energy is a result of an error cancellation
of one to two orders of magnitude.

Having assessed the accuracy of the TRD we now calculate the adsorption energy
of the water molecule on \textit{h}BN using the AVTZ pseudized Gaussian basis
set. The evaluation involves the decomposition of a Coulomb vertex with
$N_o=68$, $N_v=1564$, $N_F=0.33N_G=1450$, and $N_R=3N_F=4350$.
The computational cost to obtain the decomposed matrices is roughly 3000 CPU
hours with 256 iterations. The results of the adsorption energy are shown in
Table~\ref{tab:adsorption}. In order to grasp a physical insight of the
system, we compare the CCD results with RPA+SOSEX and MP2
calculations~\cite{paier_assessment_2012}.
MP2 theory usually overestimates dispersion driven interactions,
although in the description of BN bilayer interaction is fortuitously
accurate~\cite{hummel_many_2016}. Consequently, one expects MP2 theory to
slightly overestimate the adsorption energy, whereas RPA+SOSEX is likely to yield a very accurate estimate.
It is not surprising that CCD underbinds
the water molecule, since there exist findings that indicate the inability of CCD
for an accurate description. Higher levels of theories, such as inclusion of
the single excitations and the perturbative triples, are required for a more
appropriate treatment. Nevertheless, the purpose of the current work is to
examine the accuracy and efficiency of the newly developed methodologies
rather than the accuracy of the method itself. The CPU hours required for the
CCD calculations obtained with and without the TRD technique are summarized in
Table~\ref{tab:timing}. The time for the evaluation of the particle/particle
ladder term per iteration is as much as 43 times faster using a decomposition
with $N_R=2N_F$ and 22 times faster with $N_R=3N_F$. This constitutes a
significant gain in the computational effort of coupled cluster methods with
only slight compromise in accuracy.

\begin{table}
  \caption{\label{tab:adsorption} 
    Adsorption energies of water on \textit{h}BN obtained using the pseudized
    Gaussian basis sets at different levels of theory. RPA+SOSEX
    calculations were performed using DFT PBE orbitals as reference, whereas
    MP2 and CCD using HF ones.
  }
  \begin{ruledtabular}
    \begin{tabular}{lccc}
      Basis set &  RPA+SOSEX &  MP2  & CCD  \\ \hline
      AVDZ      &    62    & 83  & 54 \\
      AVTZ      &    72    & 92  & 62 \\
      AV(D,T)Z  &    76    & 95  & 65 \\
    \end{tabular}
  \end{ruledtabular}
\end{table}

\begin{table}
  \caption{\label{tab:timing} 
    CPU hours per iteration comparing CCD calculation with and
    without the factorized Coulomb integrals. In parenthesis we
    denote the part for evaluating the particle/particle ladder term.
  }
  \begin{ruledtabular}
    \begin{tabular}{l@{\quad\qquad}rr@{\quad\quad}rr@{\quad\quad}rr}
      Basis set & \multicolumn{2}{c}{$N_R=2N_F$} &
        \multicolumn{2}{c}{$N_R=3N_F$} & \multicolumn{2}{c}{no TRD}  \\ \hline
      AVDZ  &   39 &  (13)  &   49 & (24)  &  100 &   (75)    \\
      AVTZ  &  259 &  (28)  &  258 & (55)  & \footnote{estimation
        based on the AVDZ basis set.}1443 & (1212) \\
    \end{tabular}
  \end{ruledtabular}
\end{table}

\section{Conclusions}
\label{sec:Conclusions}
In this work we have outlined an algorithm to obtain a low rank tensor approximation of
the Coulomb integrals having the same algebraic structure as its definition
from the molecular orbitals:
\begin{equation}
\begin{array}{rcc c ccccc}
  V^{pq}_{sr} &=&
    \displaystyle\iint\limits_{{\rm d}{\bf x}\,{\rm d}{\bf x'}} &
    {\psi^\ast}^p({\bf x}) & {\psi^\ast}^q({\bf x'}) &
    \displaystyle\frac1{|{\bf r} - {\bf r'}|} &
    \psi_r({\bf x'}) & \psi_s({\bf x}) \\[3.5ex]
  &\approx&
    \displaystyle\sum_{RS} &
    {\Pi^\ast}^p_R & {\Pi^\ast}^q_S &
    {\Lambda^\ast}^F_R \Lambda^S_F &
    \Pi^S_r & \Pi^R_s\,.
\end{array}
\end{equation}
The factorization is obtained by fitting $\Lambda^S_F {\Pi^\ast}^q_S \Pi_r^S$
to auxiliary three index quantities referred to
as Coulomb vertices that are calculated from a resolution of identity approach
using a plane wave basis set. In this manner the scaling of the computational
cost for obtaining the low rank tensor approximation with respect to system size does not
exceed $\mathcal O(N^4)$.
To reduce the prefactor of the computational cost further we have outlined
an approach to further compactify the representation of the Coulomb vertices.
We linearly transform the momentum index of the Coulomb vertices into a (truncated) basis
referred to as an optimized auxiliary field.
The accuracy of this truncation is systematically improvable using a single parameter that is used for
the truncation of a singular value decomposition.
The tensor factorization of the transformed Coulomb vertices is achieved using a regularized
alternating least squares algorithm that converges rapidly using about $10^2$ iterations only.
In contrast, the nonregularized alternating least squares algorithm would require $10^5$--$10^6$ iterations.
We stress that we employ no prior assumptions for the real space grids used for expanding the 
low order tensors.

Once obtained, the tensor factorization of the Coulomb integrals can be employed to reduce the scaling of
the computational cost of distinguishable coupled cluster theory to $\mathcal O(N^5)$ without further
approximations. We demonstrate that the factorization can also be used to reduce the
computational cost for evaluating the computationally most expensive term (particle/particle ladder diagram)
in the coupled cluster doubles amplitude equations for the case of water adsorption on the \textit{h}BN
monolayer system. 
For system sizes containing 136 electrons in 1632 orbitals we achieve substantial reductions in the
computational cost that are on the order of a factor 10--20 without compromising the accuracy and introducing
any further approximation.

Future work will focus on combining the outlined techniques with explicitly correlated methods and
finite size corrections in order to significantly expand the scope of periodic coupled cluster
theory calculations using plane wave basis sets for solid state systems~\cite{Gruneis2015,Liao2016}.


\section*{Acknowledgments}
We want to acknowledge fruitful discussions with Edgar Solomonik and
Alexander Auer.


\bibliography{TRD}

\end{document}